\documentclass[pra,twocolumn,showpacs,superscriptaddress,10pt,nofootinbib]{revtex4} 
\usepackage{amsmath,amssymb,graphicx}

\usepackage{color,times}

\definecolor{darkblue}{rgb}{0, 0, 0.8}
\usepackage[colorlinks=true, breaklinks=true, linkcolor=darkblue, citecolor=darkblue, urlcolor=darkblue]{hyperref}
\newcommand{\ket}[1]{|#1\rangle}
\newcommand{\bs}{\boldsymbol}

\begin{document}

\title{Single-Atom Addressing in Microtraps for Quantum-State Engineering using Rydberg Atoms}

\author{Henning Labuhn, Sylvain Ravets, Daniel Barredo, Lucas B\'eguin, Florence Nogrette, Thierry Lahaye, and Antoine Browaeys}

\affiliation{Laboratoire Charles Fabry, UMR 8501, Institut d'Optique, CNRS, Univ Paris Sud 11,\\
2 avenue Augustin Fresnel, 91127 Palaiseau cedex, France }


\begin{abstract}
We report on the selective addressing of an individual atom in a pair of single-atom microtraps separated by~$3\;\mu$m. Using a tunable light-shift, we render the selected atom off-resonant with a global Rydberg excitation laser which is resonant with the other atom, making it possible to selectively block this atom from being excited to the Rydberg state. Furthermore we demonstrate the controlled manipulation of a two-atom entangled state by using the addressing beam to induce a phase shift onto one component of the wave function of the system, transferring it to a dark state for the Rydberg excitation light. Our results are an important step towards implementing quantum information processing and quantum simulation with large arrays of Rydberg atoms. 
\end{abstract}

\maketitle 

Cold neutral atoms are a promising platform for quantum computation and quantum simulation~\cite{bloch2012}. Their weak interactions in the ground state lead to long coherence times. Using highly excited Rydberg states allows one to switch on and off the strong interactions that are necessary for engineering many-body quantum states~\cite{saffmanRMP}. For many of those experiments it is desirable to confine single atoms at well-defined positions separated by a few $\mu$m, which can be achieved e.g. using arrays of optical tweezers~\cite{nogrette2014}. Another requirement is the selective manipulation of individual atoms in the ensemble. This can be done by applying static field gradients, or a laser beam focused to one single trap site, which induces a frequency shift at the targeted site. Such techniques have been demonstrated with trapped ions~\cite{naegerl1999,haeffner2005,monroe2013} and neutral atoms in optical lattices~\cite{dumke2002,saffman2004,karski2010,weitenberg2011,schlosser2011,fukuhara2013}.

\begin{figure}[t]
\centering
\includegraphics[width=86mm]{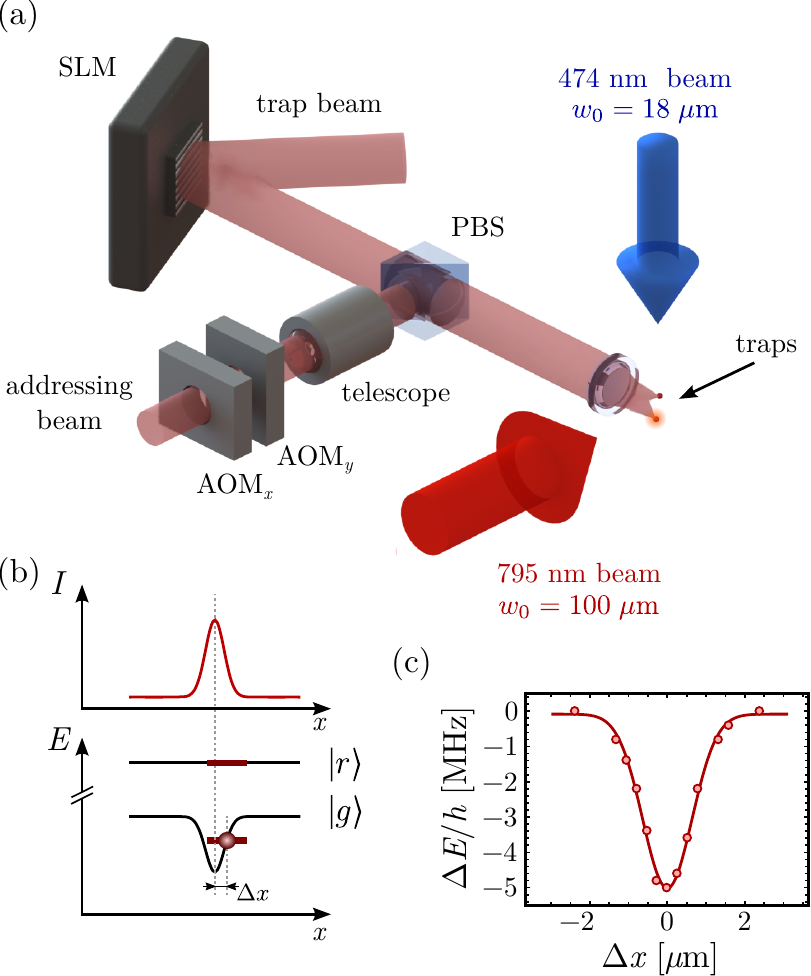}
\caption{(color online) (a)~Sketch of the experimental setup. The two microtraps are created by a red-detuned 850-nm laser beam on which an appropriate phase is imprinted using a spatial light modulator (SLM), and focused by a high-NA aspheric lens in a MOT. The addressing beam is superimposed to the trap beam by a polarizing beamsplitter cube (PBS), and focused down on the targeted atom by the same aspheric lens. The two perpendicular AOMs can be used for precise dynamical $x{\mathrm -}y$ positioning of the addressing beam. A telescope is used to conjugate the AOM plane with the aspheric lens, to avoid clipping when the addressing beam is stirred. (b)~The light-shift $\Delta E$ of the ground state of the targeted single atom is directly proportional to the intensity $I$ of the addressing beam at the position $x$ of the atom. (c)~Measured light-shift $\Delta E$ as a function of the distance $\Delta x$ between the addressing beam and the targeted trap, yielding a $1/e^2$ radius of the addressing beam of $w_0 \simeq 1.3~\mu$m.}
\label{fig:setup}
\end{figure}

In previous work~\cite{beguin2013,barredo2014}, we have demonstrated quantum-state engineering with single atoms held in two and three optical microtraps, by using the Rydberg blockade mechanism with global excitation of the atoms. Extending these studies to a larger number of atoms and to wider classes of quantum states requires extra tools. A first step towards this goal was our recent demonstration of single-atom trapping in large arrays of optical microtraps with arbitrary geometries~\cite{nogrette2014}. Combined with global excitation, this already opens the possibility to generate interesting multi-atom entangled states, such as the W state $\ket{{\rm W}} = (\ket{rgg \ldots g}+\ket{grg \ldots g} + \ldots + \ket{ggg \ldots r}) / \sqrt{N}$, where $\ket{g}$ ($\ket{r}$) correspond to the ground (Rydberg) state. However, single-site addressing is needed to engineer other classes of quantum states. For instance, the realization of the collective CNOT-gate of Ref.~\cite{mueller2009} that can be used to create the state $\ket{{\rm GHZ}} =  (\ket{gg \ldots g} + \ket{rr \ldots r}) / \sqrt{2}$, requires to single out one control atom whose state determines the state of the remaining, target atoms.

Here we demonstrate the selective addressing of one single $^{87}$Rb atom among two atoms held in microtraps separated by 3~$\mu$m, by shining a tightly-focused, red-detuned 850~nm laser beam on it. This addressing beam induces a frequency shift on the ground state of the atom, while leaving its Rydberg states nearly unaffected. This differential light-shift thus makes the addressed atom off-resonant with the Rydberg excitation laser, which is resonant for the other atom. This article is organized as follows. We first briefly describe the implementation of the addressing beam, and characterize its size and depth \emph{in situ} using a single atom. We then perform a global Rydberg excitation in the presence of the addressing beam, and observe nearly perfect suppression of excitations for the addressed atom. Finally, we use the addressing beam to perform a controlled local operation on one atom, coherently transferring the symmetric entangled state $(\ket{rg} + \ket{gr})/\sqrt{2}$ to the antisymmetric, dark state $(\ket{rg} - \ket{gr})/\sqrt{2}$.

Our experimental setup, schematically shown in Fig.~\ref{fig:setup}(a), has been described previously~\cite{nogrette2014,beguin2013,barredo2014}. We use a Spatial Light Modulator~(SLM) to create two microtraps, separated by a distance of $3\;\mu {\rm m}$ in the focal plane of a high-numerical aperture aspherical lens. The traps, each with a $1/e^2$-radius of about $1\;\mu{\rm m}$ and a depth of $U_0 \approx h \times 20$~MHz, are focused in $^{87}$Rb MOT. Due to fast light-assisted collisions, we only trap either zero or one atom per trap~\cite{sortais2007}, and trigger the experiment on the presence of one atom in each trap. The temperature of the atoms in the traps is approximately $50\;\mu$K. We coherently couple the ground state $\ket{g} = \ket{5S_{1/2},F=2,m_f=2}$ to the Rydberg state $\ket{r} = \ket{nD_{3/2},m_j = 3/2}$ (with $n$ in the range 50--100) via a two-photon transition, with the wavelength of the excitation lasers being 795~nm and 474~nm. During the excitation, of duration $\tau$, the traps are switched off. The detuning from the intermediate state $\ket{5P_{1/2},F=2,m_f=2}$ is $2 \pi \times 740$~MHz. After the excitation pulse, we measure the states of both atoms. Repeating the experiment for about 100 times, we reconstruct the populations $P_{ij}$ of the two-atom states $\ket{ij}$, where $i,j$ can take the values $g$ and $r$.

The $1/e^2$ radii of the lasers used for Rydberg excitation are 100~$\mu$m for the 795-nm beam, and 18~$\mu$m for the 474-nm beam. This configuration prevents the direct addressing of a single trap. To achieve single-site addressability, we thus superimpose a second 850~nm laser beam onto the trapping beam, which induces a light-shift proportional to the intensity on the ground state of the atom at the targeted site. Orthogonal polarizations and a frequency difference of about $200$~MHz prevent interference between the trapping and addressing beams. The addressing beam has a $1/e^2$-radius of $w_0\simeq 1.3\;\mu$m in the focus, slightly larger than the trap size. This choice results from a trade-off between two opposite requirements, namely minimizing alignment sensitivity and inhomogeneous light-shifts (which favors a large $w_0$) and minimizing cross-talk (which implies choosing a small $w_0$). For a perfectly Gaussian beam with $w_0\simeq1.3\;\mu{\rm m}$, one expects theoretically that if one atom is addressed by a light-shift of 10~MHz, the second atom 3~$\mu$m away experiences a light-shift of only 0.2~kHz, which is negligible as compared to the other relevant frequencies in the experiment. An electro-optic modulator enables fast (about 10~ns) switching of the addressing beam. In addition, two AOMs can be used for dynamical $x-y$ positioning of the addressing beam with respect to the targeted trap.

In a first experiment, we measure the intensity profile of the addressing beam \emph{in situ} by performing Rydberg spectroscopy on a single atom. For different positions $\Delta x$ of the addressing beam with respect to the targeted atom, we scan the frequency of the Rydberg excitation lasers. As mainly the ground state experiences a light-shift $\Delta E$ proportional to the addressing beam intensity, the resonance frequency for Rydberg excitation is shifted by $\Delta E$ [see Fig. \ref{fig:setup}(b)]. Figure \ref{fig:setup}(c) shows the measured light-shift as a function of $\Delta x$. A Gaussian fit gives a $1/e^2$ radius $w_0 = 1.3\pm0.1~\mu$m. The residual light-shift experienced by the nearby atom 3~$\mu$m away is below the resolution of our experiment. 

We observe that for large light-shifts, the probability to lose an atom during the sequence increases. We attribute this effect to the following: due to the finite temperature, the atom never sits exactly at the intensity maximum of the addressing beam. The fast switching on and off of the addressing beam thus imparts kinetic energy to the atom, all the more that the intensity is high. For large enough intensities in the addressing beam, this effect thus increases the probability for the atom to leave the trapping region during the experiment. However, for light-shifts below 40~MHz, this loss probability remains below 1~\%, and is thus negligible. 

\begin{figure}[t]
\includegraphics[width=80mm]{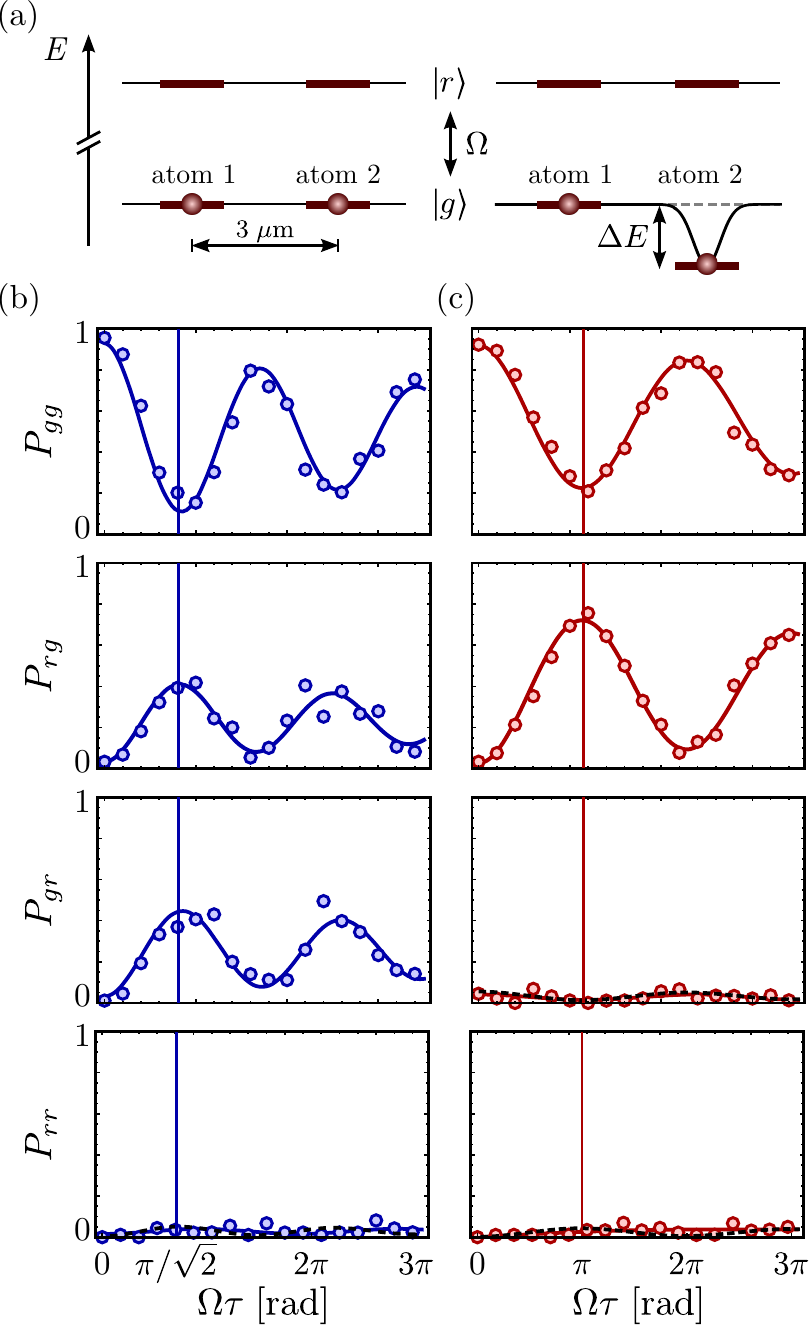}
\caption{(Color online) (a)~Two atoms, separated by 3~$\mu$m, are illuminated by light that resonantly couples the ground state $\ket{g}$ to $\ket{r}=\ket{59D_{3/2}}$ with the single-atom Rabi frequency $\Omega$. The time evolution of the populations of the two-atom states $\ket{gg}$, $\ket{gr}$, $\ket{rg}$ and $\ket{rr}$ are shown, (b)~without any addressing and (c)~with atom 2 addressed with a light-shift of $\Delta E \simeq h \times 10$~MHz. Solid lines are fits by damped sines. The vertical solid lines mark the pulse areas $\Omega\tau$ corresponding to a $\pi$-pulse for the non-addressed case (blue) and the addressed case (red).  The black dashed lines show the expected measured populations for a perfect blockade of atom 2, taking into account state-detection errors.}
\label{fig:2atoms_addressed}
\end{figure}

We now perform a Rydberg blockade experiment with two single atoms in order to demonstrate single-site addressability (Fig.~\ref{fig:2atoms_addressed}). In Ref.~\cite{urban2009}, site-resolving excitation beams were used to demonstrate blockade with two atoms separated by 10~$\mu$m. Here, we use a global excitation scheme in combination with the addressing beam, and obtain similar results, albeit with a distance between the atoms of only 3~$\mu$m. For both atoms, the ground state $\ket{g}$ is coupled to the Rydberg state $\ket{r}=\ket{59D_{3/2}}$ with a Rabi frequency $\Omega \simeq 2\pi \times1$~MHz (Fig. \ref{fig:2atoms_addressed}(a)). If the atoms were independent, they would both undergo Rabi oscillations between $\ket{g}$ and $\ket{r}$ with the Rabi frequency $\Omega$. The strong dipole-dipole interaction $U_{\rm dd}$ between the Rydberg states forbids a double excitation of the atoms if $U_{\rm dd} \gg \hbar \Omega$. This condition is largely fulfilled for the parameters chosen here: the interaction energy of two atoms in $\ket{59D_{3/2}}$, separated by $3\; \mu$m, is approximately $h\times 300$~MHz. We thus only excite the superposition state $\ket{s} = (\ket{rg} + e^{i\bs k \cdot\bs r}\ket{gr})/\sqrt{2}$, whose coupling to the two-atom ground state $\ket{gg}$ is $\sqrt{2} \Omega$ \cite{urban2009, gaetan2009} (here, ${\bs k}$ is the vector sum of the wavevectors of the excitation lasers, and ${\bs r}$ is the position of atom~$2$ with respect to atom $1$). This results in $P_{rg}$ and $P_{gr}$ oscillating between $0$ and $1/2$ with a frequency $\sqrt{2}\Omega$, as can be seen in Fig. \ref{fig:2atoms_addressed}(b). Another signature of the blockade is the suppression of double excitation $P_{rr} \simeq 0 $ (see bottom panel in Fig \ref{fig:2atoms_addressed}(b)).

If we shine the addressing beam on atom 2, we observe a strong suppression of the excitation probability for the states $\ket{gr}$ and $\ket{rr}$ (see Fig. \ref{fig:2atoms_addressed}(c)), as atom 2 is never excited to the Rydberg state $\ket{r}$. At the same time, atom 1 shows Rabi oscillations between $\ket{g}$ and $\ket{r}$ with the single-atom Rabi frequency $\Omega$. The small residual excitation probability of atom 2 that we observe is fully accounted for by the errors in our state detection~\cite{barredo2014}, meaning that cross-talk between the two traps is negligible.

\begin{figure}[t]
\includegraphics[width=70mm]{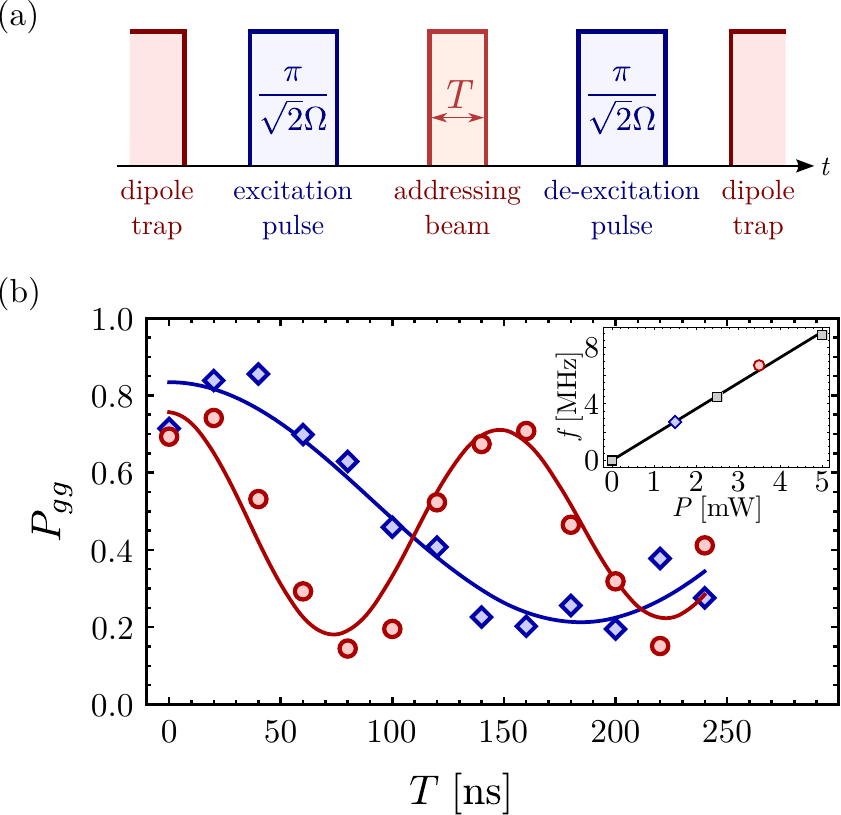}
\caption{(Color online) 
(a)~Pulse sequence for the phase manipulation: while the dipole trap is switched off, the atoms are excited to the state $\ket{s} = \left(\ket{gr}+e^{i{\bs k}\cdot{\bs r}}\ket{rg} \right)/\sqrt{2}$.  The addressing beam induces a light-shift $\Delta E$ on the ground state of atom~2, thus changing the relative phase evolution between $\ket{gr}$ and $\ket{rg}$. This is followed by a global de-excitation pulse. (b)~Population of the two-atom ground state $\ket{gg}$ after the de-excitation pulse, as a function of the addressing pulse length $T$, for a laser power in the addressing beam $P=1.5$~mW (blue diamonds) and $P=3.5$~mW (red circles). Solid lines are fits by damped sine of frequency $f$. Inset: oscillation frequency $f$ as a function of the power $P$ of the addressing beam, showing the expected linear dependence. For this experiment we use the Rydberg state $\ket{82D_{3/2}}$.}
\label{fig:phase-shift}
\end{figure}

Finally, we show that we can also use the addressing beam to directly manipulate a two-atom quantum state. Without any addressing, the excitation to the state $\ket{rr}$ is completely suppressed in the Rydberg blockade regime ($U_{\rm dd} \gg \hbar \Omega$). By applying an excitation pulse of duration $\pi / (\sqrt{2} \Omega)$ we thus prepare the atoms in the state $\ket{\psi (0)} = (\ket{gr} + e^{i\bs k \cdot\bs r}\ket{rg})/\sqrt{2}$. We then illuminate atom 2 with the addressing beam (Fig.~\ref{fig:phase-shift}(a)).  Its energy is shifted by $\Delta E$ when in the ground state, while its Rydberg state remains unaffected (see Fig. \ref{fig:2atoms_addressed}(a)).  After a time $T$ the state of the system has therefore evolved to 
\begin{equation}
\ket{\psi(T)} = \frac{1}{\sqrt{2}} (\ket{gr} +e^{-i \Delta E \, T /\hbar} e^{i\bs k \cdot\bs r} \ket{rg}).
\end{equation} 
The antisymmetric dark state $\ket{\psi(T_{\pi})} = (\ket{gr} - e^{i\bs k \cdot\bs r} \ket{rg})/\sqrt{2}$ (with $T_{\pi}=\pi \hbar/ \Delta E$) is not coupled to the ground state $\ket{gg}$. The probability to deexcite the atoms to $\ket{gg}$ is thus expected to oscillate between 0 and 1 with a frequency $f = \Delta E / h$.

Figure~\ref{fig:phase-shift}(b) shows the probability $P_{gg}$ to de-excite the atoms back to $\ket{gg}$ versus the duration $T$ of the addressing pulse. We observe the expected oscillation of the final ground state population $P_{gg}$. Due to the finite Rydberg excitation efficiency (about 90\% for our parameters), we measure a contrast of the oscillations that is lower than~1. In addition, the finite temperature of the atoms in the experiment leads to a small motion of the atoms during the sequence, implying that (i) the phase ${\bs k}\cdot{\bs r}$ imprinted by the excitation pulse is not exactly canceled out by the de-excitation pulse~\cite{wilk2010}; and (ii) the light shift $\Delta E$ seen by atom 2 fluctuates from shot to shot. Averaged over many runs, both effects lead to a decreased contrast and a finite damping of the observed oscillations. To take these effects into account, we fit the data with a damped sine of the form $P_{gg}(T) = A + B \exp(- \gamma t) \cos(2\pi f  T)$, with the oscillation frequency $f$ and the damping rate $\gamma$ as adjustable parameters. Repeating the experiment for different powers of the addressing beam, we obtain the expected linear dependance of $f$ with the applied light shift on the atom [see inset of Fig.\ref{fig:phase-shift}(b)]. This demonstrates our ability to perform some controlled local operations on qubits in a quantum register.  

In conclusion, we have shown that we can selectively prevent one single atom in a pair of single-atom traps from being resonant with Rydberg excitation lasers, with no measurable cross-talk with a neighboring atom as close as 3~$\mu$m. We also demonstrated the use of the addressing beam to perform a local operation in a system of two atoms.  Our scheme is easily scalable to a larger number of traps. These techniques will prove useful for a variety of applications in quantum simulation and quantum information processing with Rydberg atoms. For instance, they open the possibility to selectively address a single qubit in a larger ensemble, e.g. as a control atom for realizing collective quantum gates~\cite{mueller2009}, or to excite a single atom to a different Rydberg state, allowing to study the transfer of excitations along a Rydberg chain~\cite{wuester2010}.

We thank Yvan Sortais for helpful advice about the optical design. We acknowledge financial support by the EU [ERC Stg Grant ARENA, AQUTE Integrating Project, FET-Open Xtrack Project HAIRS, and EU Marie-Curie Program ITN COHERENCE FP7-PEOPLE-2010-ITN-265031 (H. L.)], by DGA (L.~B.), and by R\'egion \^Ile-de-France (LUMAT and Triangle de la Physique, LAGON Project).

\end{document}